\documentclass{article}
\usepackage{graphicx} 
\usepackage{amsmath}
\usepackage{amssymb} 
\usepackage{natbib}
\usepackage{tabularx}
\usepackage{booktabs} 
\usepackage{array}
\usepackage{makecell}
\usepackage[utf8]{inputenc}
\usepackage[table]{xcolor}
\usepackage{geometry} 
\usepackage{enumitem}

\usepackage[colorlinks=true, linkcolor=blue, citecolor=blue, urlcolor=blue]{hyperref}

\title{A parallel monetary system based on the redeemable self-decaying money --- The ultimate hedge and safe haven of private wealth in the rising wave of over issuance of fiat and token money/stablecoin
}
\author{
    Boliang Lin\textsuperscript{a,c,*},
    Ruixi Lin\textsuperscript{b}
    \\
    \small \textsuperscript{a}School of Traffic and Transportation, Beijing Jiaotong University, Beijing 100044, China \\
    \small \textsuperscript{b}School of Computing, National University of Singapore, Singapore 117416, Singapore \\
    \small \textsuperscript{c}Beijing Laboratory of National Economic Security Early-warning Engineering, Beijing 100044, China
}
\date{}
\begin{document}
\maketitle

 \noindent \textbf{Abstract:}  A currency with stable purchasing power can always provide a psychological haven for people around the world. However, since the collapse of the Bretton Woods system, issuing more cheap currencies has become a common trend in the international community, and the legalization and over issuance of stablecoins will strengthen this trend. In this context, our study focused on a parallel monetary system based on a redeemable self-decay/devalued money(RSDM). Firstly, we point out the idea of redeeming gold at a fixed denomination with gold certificates is similar to an impossible perpetual motion machine. Only when the face value of a gold token self-decays or self-depreciates and the weight of the reduced value can compensate for the storage cost of physical gold, can it be convertible or redeemable. Secondly, we pointed out that as a modern ``good money'' under the Internet environment, it must have two basic functions: long-term value storage and zero logistics cost of money circulation. Thirdly, we found that a single type of money is difficult to shoulder the responsibility of modern ``good money''. Only a parallel monetary system, including RSDM, such as a triple-monetary system consisting of RSDM, domestic fiat and major international reserve currencies, can form the ultimate safe haven of wealth and safeguard the reverse Gresham law. Based on this analysis, we build an integer programming model for currency optimization selection in a multi-monetary pool. Fourthly, several potential application scenarios of RSDM in the real world were discussed, including a new approach to activate dormant gold assets in India based on RSDM, and the gold monetization scheme in the United States. Finally, the demand for RSDM with precious metals as collateral was analyzed, providing theoretical support for establishing a sound parallel monetary system based on RSDM.

 
\section{Introduction}

Why did all the redeemable commodity money tokens break their promises that can be converted into mortgage assets at any time in the end? for example, the `Jiaozi' appeared in China about a thousand years ago \citep{battilossi2020}, early banknotes in Europe, the US dollar during the Bretton Woods system, etc. The real reason hidden in the back of the depreciation of banknotes is the black hole of storage charge of the mortgage assets overtime after commodity money tokenization \citep{lin2021}. It is the most basic attribute of a ``good money'' that has the function of storing value and can achieve efficient and low-cost circulation, which means it have low logistics costs (circulation and warehouse). Therefore, a currency with stable purchasing power can always provide a psychological haven for people around the world. As is well known, the essence of fiat currency is a non-convertible receipts issued by the government, and its natural flaw is indiscriminate issuance \citep{paul2008}. Any person or regime with the power to issue currency often does not miss the opportunity to legally occupy people's wealth so easily.

Relatively speaking, due to resource constraints, it is difficult to indiscriminately issue precious metal coins. Although commodity money such as gold coin generally have better stability compared to fiat currencies, we should also recognize that due to the asynchronous changes in production costs of various commodities, it is difficult to maintain a constant relative price between any two commodities (such as gold and grain). Obviously, the periodic instability of resource prices makes it difficult for the government to quickly increase or decrease the supply of commodity money, resulting in unstable prices. From this perspective, compared to fiat currencies with good issuance discipline, a single gold standard does not have significant advantages.

Considering that money is a social custom, a valuable numerical concept, and a unit of account. Therefore, the most ideal monetary system should be a symbolic currency that can maintain stable purchasing power. For example, a fiat/credit currency with good monetary discipline. However, throughout the history of currency, there have been few successful cases of effectively constraining the issuance of credit currency, or in other words, a well-managed fiat system is rare. It is not difficult to find that in various historical stages, there have been many cases of excessive issuance of various credit currencies leading to malignant inflation, while there have been very few cases of overnight significant depreciation of commodity money such as gold and silver coins. Therefore, there was once a strong call to return to the gold standard (or commodity standard), as it had a certain user base. Considering that the quantity of resources as commodity money is difficult to synchronize with GDP and their logistics costs are high, it is difficult for a single commodity money to shoulder the responsibility of modern currency. Therefore, only a parallel currency system composed of commodities with relatively stable mining costs (such as gold, silver, copper, and other monetary metals) and reputable international reserve currencies is suitable as the cornerstone of a price stability system.

Although virtual currencies such as Bitcoin have the advantages of fast cross-border payments, minute level trade settlements, high transaction efficiency and low transaction fees, their value is extremely unstable compared to major fiat currencies (such as the US dollar), and there are significant risks for users to use them for long-term contract settlements. From the current perspective, the role of Bitcoin is more of an asset attribute rather than a monetary attribute. Therefore, stablecoins pegged one-to-one with fiat such as the US dollar have emerged, allowing users to enjoy a relatively stable trading environment without leaving the blockchain. Currently, the wave of stablecoin issuance is sweeping across the globe.

Considering that stablecoins have faster transaction speeds compared to traditional settlement systems, it will improve the efficiency of currency usage. In addition, if stablecoins are used as collateral for loans or refinancing, it will promote credit expansion, so the issuance of stablecoins will inevitably lead to credit creation and generate a large money multiplier. What is particularly alarming is that a potential wave of over issuance of tokens such as stablecoins/RWAs (Real World Assets) could become an invisible driver of global inflation. Since the collapse of the Bretton Woods system, issuing more cheap currencies has become a common trend in the international community, and obviously, the legalization of stablecoins will strengthen this trend.

The future trend of currency issuance is likely to be that the supply of international liquidity is difficult to control, and inflation and foam economy are more frequent. When the wealth stored in paper currency may evaporate overnight, will we still feel at ease using these banknotes as a way to store wealth? When commercial banks or financial institutions can simply create “electronic currency” for credit expansion, can these currencies still serve as a stable measure of value? As there are more and more types and volumes of digital currencies without commodity anchors, the increase in consumer goods cannot keep up with the increase in currency. How can people hold onto their wealth? The answer is: Honest tokens with redeemable collateral can build the ultimate moat of public wealth.

In history, the main characteristics of commodities as “general equivalents” include: stable physical properties, scarcity of resources, global acceptance, non-credit, non-sovereign properties (not dependent on the credit and commitment of any country or institution), and the logistics cost during the transaction settlement is very low. Therefore, the modern honest money in the Internet environment should have the store of value function of precious metals and the advantage of near zero logistics costs of digital currencies. History and reality tell us that traditional fiat and metal tokens, as well as emerging virtual currencies such as stablecoins, cannot achieve such dual advantages. However, the RSDM (a Redeemable Self- Decay/Devalued money) that will be discussed in this paper is precisely such a modern honest money with the double attributes of store of value and near zero logistics cost. It seems that using RSDM as a temporary replacement money is a good choice for people to hold onto their wealth. Macroscopically speaking, RSDM, as a supervisory money to prevent the excessive issuance of anchor free currencies, can build the ultimate moat of public wealth.

The remainder of this paper is organized as follows. Section 2 presents a comprehensive literature review on related studies. Section 3 point out that the idea of redeem metal with value symbols at any time is an impossible perpetual motion machine. We develop a self-devalued commodity money token that can be redeemed for gold in Section 4. A parallel monetary system based on a redeemable self-decay/devalued money(RSDM) is studied in Section 5. Section 6, we build an integer programming model for currency optimization selection in a multi-monetary pool. Section 7, several potential application scenarios of RSDM in the real world were discussed, including a new approach to activate dormant gold assets in India based on RSDM. In section 7, the demand for RSDM with precious metals as collateral was analyzed. Finally, conclusions and research prospects are drawn in Section 9.

\section{Related Work}
Paper money, in effect, comes to redeem the metal upon which it was originally based, i.e. the paper currency is actually representative paper that redeems metal \citep{gatch1996}. In fact, \citet{kreigel2021} has found that as the safety and transport of gold represented a substantial cost and mechanisms came in to play to substitute commodity money with bills of exchange and time contracts such as futures and options. And the public accepted the valueless paper notes because they had faith in the fiduciary holding of ``real'' commodity money in the banks' reserves. An early research work that proposed the concept of decaying currency is \citet{lin2016}. The work reveals the attenuation mechanism of anchor of the commodity money from the perspective of logistics warehousing costs, such as the cost of safe storage and transportation of precious metals, and propose a novel Decayed Commodity Money (DCM) for the store of value across time and space. In addition, this study suggests that the DCM can also avoid the defects that precious metal money is hoarded by market and credit currency often leads to excessive liquidity. On this basis, \citet{lin2021} provide a clear review of the evolution history of currency from the perspective of logistics costs, and emphasizes that the key functions of money are the store of value and low logistics (circulation and warehouse) cost. Although commodity money (such as gold and silver) has the advantages of a wealth store, its disadvantage is the high logistics cost. In comparison to commodity money, credit currency and digital currency cannot protect wealth from loss over a long period while their logistics costs are negligible. They further proved that there is not such honest money from the perspective of logistics costs, which is both the store of value like precious metal and without logistics costs in circulation like digital currency. 

Overall, the limited adjustability of the gold supply makes the gold standard ill-suited to accommodate changes in GDP, and a complete reversion to a metallic standard lacks practical feasibility. Future monetary system reform should aim to combine the stability advantages of the gold standard with the flexibility of modern monetary policy.

Regarding the history of the gold standard, \citet{cooper1982} has provided a relatively systematic review, especially on the situation before the 1980s. This paper pointed out that the main interest in reviving gold lies primarily in a desire to eliminate inflation and preserve a noninflationary environment. The historical analyses of the gold standard, the mechanics of the gold standard, the feasibility of returning to the gold standard, and ``digital gold'' in the form of cryptocurrencies are also extensively discussed in the collection of essays ``The Gold Standard: Retrospect and Prospect'' \citep{newman2021}. Although few academic economists today endorse a gold standard, historical data show that actual gold standards have outperformed actual fiat standards in at least some respects. Gold standards have exhibited: (1) lower mean inflation rate; (2) lower price level uncertainty, hence deeper long-term bond markets; and (3) greater fiscal discipline \citep{white2015}, etc. \citet{bordo1995} argued that the gold standard, as a contingent commitment mechanism, constrained the monetary authority from altering planned future policies, thereby enhancing the credibility of money. \citet{bazot2022} found that, under the classical gold standard with high capital mobility and fixed-exchange rates, central banks mitigated the impact of international shocks on domestic interest rates by adjusting the size of their balance sheets; this highlights the institutional advantage of the gold-standard framework, where international constraints were partly offset by balance-sheet policies. \citet{fernandez2023} present a micro-founded monetary model of a small open economy to examine the behavior of money, prices, and output under the gold standard. The gold standard ensures long-term price stability as the quantity of money and prices only temporarily deviate from their steady-state levels. \citet{dimartino2025} showed that the fixed parity under the gold standard, sustained by discount rate policy, lowered exchange rate volatility and thereby promoted international trade. \citet{mitchener2016} highlighted that the gold-standard peg promoted external trade by lowering currency risk, though adverse export-price shocks could still raise the chance of a devaluation.

However, a broad consensus holds that the rigidity of the gold standard limits the central banks' ability to respond to economic crises. \citet{bordo2009} compared the crises of 1929 and 2008 and conclude that the modern central bank toolkit, such as quantitative easing, provides policy space to address systemic financial risks that the gold standard lacks. \citet{mitchener2015} confirmed that countries which left the gold standard earlier and imposed capital controls recovered significantly faster than those that remained on gold. \citet{obstfeld2017} noted that, compared with the Bretton Woods system and floating exchange-rate regimes, the gold standard displays marked institutional rigidity when confronting financial crises. \citet{ho2022}, through a counterfactual analysis of Germany from 1930 to 1932, found that adherence to the fixed Reichsmark parity under the gold standard deepened the recession. For emerging-market economies, the credibility of gold standard commitments was often limited, as documented by \citet{mitchener2015b}. \citet{pensieroso2024} pointed out that the gold standard prolonged both the depth and the duration of the Great Depression. \citet{kramer2022} study how political institutions affected the decision of countries to adhere to the classical gold standard. And they find that the probability of adherence to the gold standard before World War I was ceteris paribus lower for countries which were more democratic. \citet{valadkhani2022} consider a novel approach to examine when gold may be an effective inflationary hedge, and the conditions under which this relationship holds. They found that when inflation is high, gold exhibits significant responses to changes inflation. However, when inflation is moderate or low, gold remains somewhat non-responsive. They thus argue that such asymmetric and size-dependent responses are the main causes of the lack of consensus in the literature regarding the hedging capability of gold.

There is also a lot of research work on returning to the gold standard. For example, \citet{cutsinger2020} discussed the prospect of returning to such a monetary system raises several important questions that would need to be addressed prior to its implementation. And he/she believes that returning to the gold standard would be feasible in the technical sense. \citet{yaacob2014}'s study provide evidence that the gold standard era is the most stable era thus justifies the return to gold currency. Their results show that the rate of inflation and the value of world gold are much lower and more stable during the gold standard phases than the fiat money. This indicates that the move to return to gold currency is more apt in the bid to ensure global economic stability. However, more efforts are needed to resolve the implementation mechanism. One of the paths suggested by \citet{white2012} for transitioning to a new gold standard is to let a parallel gold standard grow up alongside the current fiat dollar.

As is well known, the natural extension of the gold standard is the commodity money standard. \citet{cesarano2014} point out That monetary arrangements had been based on metallic money almost uninterruptedly for twenty-five hundred years is no historical accident but rather the result of an equilibrium process propelled by the optimal selection of media of exchange under the constraint of technology. \citet{selgin2014} proposed the concept of synthetic commodity money, and believed that it may supply the foundation for a monetary regime that does not require oversight by any monetary authority, yet is able to provide for a high degree of macroeconomic stability. \citet{rolnick1998} examines the behavior of money, inflation, and output under fiat and commodity standards. They pointed out that by a monetary standard, we mean the objects that serve as the unit of account and that back the objects that circulate as generally accepted means of payment. Under a commodity standard, the unit of account is a fixed amount of the commodity. \citet{sussman2003} presents a theory of inflation in commodity money and show that anticipated stabilization reduces demand for commodity money. \citet{hendrickson2022} find that a pure commodity money regime can only generate an efficient stationary equilibrium by divine coincidence or by giving policymakers control over the supply of the commodity. The introduction of bank notes makes it much more likely that the economy will achieve an efficient equilibrium. \citet{costabile2022} argue that the system of pure commodity money is the cornerstone of Ricardo's monetary theory because it equalizes the value of money to the natural value of gold. In commodity money systems, the price level is equal to the relative price of gold for ``commodities in general'', and money is endogenous. Commodity money is a specific, complex institutional system.

The “Digital Gold Standard”, emerging from the convergence of blockchain technology and classical monetary theory, has become a focal point in recent academic and policy debates. In fact, besides stablecoins with precious metals as collateral assets, the tokenization of real-world assets (RWAs) may also become a key bridge connecting traditional commodity markets with the digital financial ecosystem. In the face of a scholarly consensus that all currency is `token' or `fiat' in nature, \citet{vasantkumar2019} seeks to make sense of the popular persistence of `commodity' or `metallist' understandings of money's value and explored that the semiotic convergence between gold, Bitcoin and modern paper money. \citet{bordo2020} proposed that an algorithmic anchoring of central bank digital currencies (CBDCs) to gold could achieve an ``elastic gold standard.'' This approach would dynamically adjust the gold-backing ratio through smart contracts, balancing monetary discipline with economic growth needs. \citet{auer2022} further investigate stablecoin collateral design, advocating layered or hybrid collateral frameworks (potentially including traditional assets and reserves) to optimize stability-efficiency trade-offs. Technological implementation paths vary significantly. PAX Gold (PAXG) and Tether Gold (XAUT) exemplify token models backed by 1:1 physical gold reserves. The World Bank \citep{worldbank2020} highlighted gold's role in central bank reserves, noting that countries increasing gold holdings during high inflation periods more effectively curb domestic inflation—a phenomenon aligning with mechanisms of gold-anchored stablecoins.

Therefore, this paper aims to make the following contributions to existing literature:

A. We propose a parallel monetary system based on a redeemable self-decay money, aiming to provide people with a safe haven for wealth amid the wave of over issuance of fiat currencies and tokens/stablecoins.

B. In order to establish a sound parallel monetary system, we built an optimization model based on integer programming to solve how to select currency types from monetary pool.

C. We propose a monetization scheme for Indian dormant gold based on RSDM. 

D. We pointed out that RSDM, as the ultimate competitor with the US dollar, will become a soft constraint to prevent the over-issuance of fiat currency.

\section{The idea of redeeming metal based on the fixed denomination on paper or digital currency at any time is an impossible perpetual motion machine}
In this section, we will prove this conjecture in two situations, namely, that a 100\% redeemable token is impossible to exist. Just like how the laws of thermodynamics prove that perpetual motion machines cannot exist. For the convenience of proof, we have designed such a virtual scene: Assuming that the company LinBL issues a token (stablecoin, certificate) ‘LinAG30’ with silver as collateral on January 1, 2030, each token corresponds to one gram of silver, stored in a professional vault in Switzerland or Hong Kong. Holders can use the tools provided by LinBL to verify the silver storage information related to this token, ensuring the security and transparency of the collateral assets.
\subsection{Case 1: Tokens such as silver or gold certificates without expiration timestamps}
In this case, assuming that the token is the same as fiat currency and there is no expiration date set on the paper, it can continue to circulate and be used.

\textbf{Theorem}: There is no such token that redeem metal/collateral based on the static denomination printed on paper or digital currency at any time. 

\textbf{Proof}:

The discussion in this section largely follows the previous work of the author\citep{lin2021}. We assume that the customer $k$ buys $n_{k}$ tokens of LinAG30 at time $t_{k}$ with $n_{k}$  grams of silver (or in equivalent fiat money at the price of silver at that time, like US dollars). After a period of time, the customer or the person who is in possession of tokens redeems silver at time $t_{k}^{'}$ . The customer will pay the processing fee to the LinBL company is $\beta$  per token, including the cost of token destruction after redemption and the handover fee for physical silver, thus the gross profit issuer can obtain is formulated as follows:

\begin{equation}
C^{\text{Profit}}=\sum_{k}\beta\times n_{k}
       \label{eq:1}
\end{equation}
\hspace{2em}The logistic cost of the issuer can be formulated as follows:
\begin{equation}
C^{\text{Warehouse}}=\sum_{k}\alpha\times n_{k}\times (t_{k}^{'}-t_{k})
       \label{eq:2}
\end{equation}
where, $\alpha$  is warehouse cost of saving one gram of silver per day, which consists of transportation cost between the warehouse and the delivery location, settlement cost, storage cost, management cost, finance charge etc. It should be pointed out that it is assumed here that the fiat currency or some accounting currency used to measure logistics costs has not experience depreciated during period $\Delta=(t_{k}^{'}-t_{k}) $.

If the gross profit of issuer can't cover logistic cost, i.e., the follow condition will be met,
\begin{equation}
C^{\text{Profit}}<C^{\text{Warehouse}}
       \label{eq:3}
\end{equation}
then the issuer will go bankrupt. Obviously, there exist always such a day when condition (3) is met. Without loss of generality, assuming that the whole system has only one customer, equation (3) can be simplified as $\beta\times n < \alpha\times n\times (t^{'}-t)$. We can always wait until such a day  $t^{'}$  when the customer comes to redeem silver with the token, which makes the following equation (4) work.
\begin{equation}
t^{'}-t > \beta/\alpha
       \label{eq:4}
\end{equation}
In order not to go bankrupt, after collecting the user's collateral, the issuer often invests the collateral during the user holds the token, which will earn interest for issuer. In this case, investment risk will be generated. Although investment risks such as buying US Treasury-Bonds and depositing in banks are relatively low, risks always exist.

\subsection{Case 2: Token with limited usage period and expiration timestamp}
If the issuer limits the usage period of the token and cannot determine the average circulation time of the token in the market, the token issuing institution will tend to charge collateral storage fees based on the time difference between the purchase date and the deadline. That is, the issuer will charge storage fees for the collateral according to formula $\beta=(t^{\text{Deadline}}-t) \times \alpha$  , where $t^{\text{Deadline}}$  is the expiration date of the token. If the predicted average circulation time of a token is $\bar{t}$ , the storage fee can be charged according to formula $\beta=\bar{t} \times \alpha$ . However, accurately predicting the value of $\bar{t}$  is a difficult task.

This rule is unfair to short-term holders of those tokens. Moreover, how issuers handle expired tokens is also a difficult problem to solve. If the storage fee is set according to time, the token becomes the Decayed Commodity Money (\citet{lin2016}).

\subsection{Case 3: The issuer uses the customer's collateral assets for risky investments}
Assuming that the issuer's own reserve assets are $\mathbb{R^{\text{Assets}}}$, the customer deposits obtained from issuing tokens/stablecoins are $\mathbb{R^{\text{User}}}$, the average duration of customer holding token is $\Delta$, the investment income of the issuer during time period $\Delta$ is $\Delta^{\text{Income}}$, and the total expenditure of the issuing institution during the period is $\Delta^{\text{Expenses}}$, when condition 
$\mathbb{R^{\text{Assets}}} + \mathbb{R^{\text{User}}} + \Delta^{\text{Income}}- \Delta^{\text{Expenses}} <\mathbb{R^{\text{User}}}$
is met, that is, when $\mathbb{R^{\text{Assets}}} + \Delta^{\text{Income}} <\Delta^{\text{Expenses}}$ is satisfied, the issuer will be insolvent and some customers will not be able to redeem the collateral of tokens. Because investment returns  $\Delta^{\text{Income}}$ may be negative, especially for high-risk investments such as stocks.

Based on the above analysis, it is not difficult to conclude that a redeemable collateral token must satisfy the self-devalued denomination theorem, otherwise, all redeemable promises cannot withstand the test of time. In the same way, a 100 percent Gold Dollar cannot exist stably. That is to say, when users redeem the collateral, the fixed delivery fee in advance will lead to the issuer having the impulse to invest the user's collateral assets for additional profits. Therefore, fully convertible tokens without a face value decay mechanism can exist only as “perpetual motion machines” in the utopian financial world, which cannot exist. This is a theoretical negation of the existence of such “perpetual motion machines” by the “Second Law of Thermodynamics” in finance.
\section{A self-devalued commodity money token that combines store of value and low circulation cost}
Although commodity money (such as gold and silver coin) has the advantages of a wealth store, its disadvantage is the high logistics cost. Therefore, in this section, we will discuss a redeemable self-decay/devalued money that combines store of value and low circulation cost.
\subsection{Classical honest money——major flaws and their short-lived tokens}
There is currently no unified definition for honest money. The mainstream consensus is that honest money has a relatively stable value and intrinsic value support. This type of money itself has practical value, such as gold being used to make jewelry and silver being used for industrial purposes. Here, we define metal money represented by gold and silver as classical honest money.
The major defects of metallic money are as follows:

A. The logistics cost is high, and storage and carrying are inconvenient, especially for large-scale transactions, with high carrying and transportation costs and security risks.

B. Metal money is difficult to accurately segment in small-scale transactions. If precious metals are designated as the primary currency and base metals as the secondary currency, it is difficult to fix the price comparison between two metals (such as silver and copper) due to the asynchronous changes in mining costs of different metals. Increased transaction costs and complexity.

C. Precious metals are prone to wear and tear. In long-term use and circulation, metal money may experience changes in weight and color due to wear and oxidation, which can affect its value and acceptability.

D. Easy to be scraped by criminals, resulting in a decrease in the actual weight of money. During transactions, precise weighing is required every time, which is time-consuming and increases transaction costs.

E. Difficult to distinguish between genuine and counterfeit money and purity.

F. Precious metals are often hoarded, resulting in insufficient liquidity.

Due to the high logistics costs of metallic money (especially base metals), in order to facilitate large transactions, people invented tokens, such as the ‘Jiaozi’ appeared in China about a thousand years ago, early goldsmith’s notes and banknotes in Europe, silver draft (a form of paper money) that appeared in Shanxi Bank in China two hundred years ago, the US dollar during the Bretton Woods system, and early SDRs etc. The common feature of these tokens is that the issuers promise to exchange them for gold or silver, that is, can be convertible exchange or redeem. And redeem gold or silver based on the static/fixed denomination of the token. According to the previous discussion, as long as the storage fee for the collateral is charged in proportion to the face value of the token, and the banknote's face value does not decrease over time, the promise of full redemption is a lie. For example, as we mentioned earlier Jiaozi, when the holder wants to redeem metal coins, they need to pay a 3\% storage fee, regardless of the length of storage time. Due to the 3\% storage fee only covering the storage cost for a limited period of time, the lifecycle of this token will not be too long.

Therefore, all redeemable commodity money tokens that combines store of value and Low circulation cost have been abandoned by history.

\subsection{Modern honest money: Transferring some value of the collateral through self-decay to tokenize commodity money}

As we have analyzed above, the major defects of classic honest coins are high logistics costs, while the main drawback of credit or digital currencies is the difficulty in storing value. In order to solve this dilemma, people tried many ways to monetize various warehouse receipts, similar to today's tokens, throughout history. Due to the fact that these tokens, originating from warehouse receipts, do not take into account the accumulation effect of logistics costs over time, the denomination of token does not automatically depreciate over time, or there is no note on the token indicating the rule of reducing denomination. After a long period of time, when the token holder comes to redeem the metal coin, if the delivery fee paid is not enough to cover the storage cost of the collateral, the result is either the issuer going bankrupt or issuer breaking their promise to redeem. In fact, in the third part of this paper, we have proven that such a token is theoretically impossible to exist for a long time.

Unlike classic honest money, which have inherent value as their main highlight, modern honest money require not only retaining all the advantages of classic honest money, but also adding the near zero logistics cost advantages of paper money. As early as 2016,  \citet{lin2016} proposed a new currency called DCM (\textit{\textbf{D}ecay \textbf{C}ommodity \textbf{M}oney}), which can be used as a store of value. This type of currency has the characteristic of self-decaying value over time. Therefore, DCM has the advantages of both the commodity money which has the function of store of value and credit or virtual currencies with the near zero logistics cost. In addition, DCM can also avoid the defects that precious metal money is hoarded by market and credit currency often leads to excessive liquidity. In 2021, \citet{lin2021} discovered the black hole of logistics costs to digitize or tokenize physical commodity money. And they proved that there is not such honest money from the perspective of logistics costs, which is both the store of value like precious metal and with almost zero logistics costs in circulation like digital currency. The reason hidden in the back of the depreciation of banknotes is the black hole of storage charge of the anchor overtime after digitizing commodity money. Moreover, the author renamed DCM as SDC (an honest\textit{\textbf{S}table \textbf{D}evalued \textbf{C}urrency}). This paper will more accurately name this money as an RSDM (\textit{\textbf{R}edeemable \textbf{S}elf- \textbf{D}ecaying/Devaluing \textbf{M}oney}). Since RSDM can be redeemed based on the negative interest rate rule, we can also name it RNIRM (a Redeemable Negative Interest Rate Money). 

RSDM transfers a portion of its value by self-decaying the weight of the collateral to meet the accumulated storage cost expenses when it is redeemed. Therefore, it can be in the form of paper currency or digital currency. 

To illustrate the characteristics of RSDM, we further explain it by modifying the example in \citet{lin2021} .

Supposing a currency issuing institution (a bank or a finance company) $B$ issues a type of RSDM at time $t$ (usually in the New Year's day of a certain year such as January 1, 2035). Each RSDM contains a gram of precious metal (or some combination of commodities) $m$ as a mortgage asset. The mortgaged assets are stored in the LBMA professional vault in the Swiss Alps and are supervised by a third party. Issue size is $N_B^m(t)$. Let $M_B^{\text{Token}}(t,m,W, \theta, E, \lambda)$ denote the token of an RSDM, where $t$ is the date of issue, and $m$ stands for the mortgage assets, $W$ for the weight of collateral each RSDM, $\theta$ for the attenuation coefficient of the weight of precious metal $m$, i.e., a negative interest rate. $E$ for the expiry date of RSDM and $\lambda$ is the delivery fee rate for RSDM holder to redeem for collateral. Take gold as a case, for example, $t$ is January 1, 2035, $m$ is the gold with a purity of 99.99\%, $W = 1$ gram. A negative interest rate is $ \gamma=0.04\%$ per day, i.e., reducing residual values by 0.04\% per day. In other words, the daily attenuation coefficient is $\theta = 99.996\%$. $E$ is fifty years, i.e., the expiration date of this RSDM is before December 31, 2084. $\lambda = 0.003$, it means the redemption fee equals to 3‰ of the weight of the residual value. Company B plans to issue 2 billion RSDM tokens, i.e., $N_B^m(t) = 2 \times 10^9$.

If a customer plans to buy an RSDM from company B at time $t+\Delta t'$, the residual weight of an RSDM can be calculated as follows

\begin{equation}
W(t+\Delta t') = W \times \theta^{\Delta t'}
\label{eq:5}
\end{equation}

Let $P_{B,C}^m(t+\Delta t')$ denote the price of the mortgage asset $m$, where the price is set according to a certain credit currency $C$. Then, a customer needs to pay $P_{B,C}^m(t+\Delta t')  \times W\times \theta^{\Delta t'}$ for an RSDM. In practice, the customer can give the issuer $W(t+\Delta t') $ units of the precious metal $m$ for an RSDM. In this case, the issuing institution $B$ will check whether the gold submitted by the holder meets the purity standards, and this will incur an inspection fee of $\Delta H$. Considering that the currency $C$ is not stable, $\Delta H$ could differ at different times, and the issuing institution $B$ will have to announce $\Delta H$ in advance. Of course, customers can also purchase RSDM from other holders at prices that are spontaneously formed by the market, which can be priced in a certain fiat currency or stablecoin or other types of RSDM etc.

When an RSDM holder needs to redeem the collateral at time $t+\Delta t''$, the quantity of precious metal available to the customer is

\begin{equation}
W(t+\Delta t'') = \theta^{\Delta t''}\times W-\lambda \times \theta^{\Delta t''}\times W=(1-\lambda )\theta^{\Delta t''} W
\label{eq:6}
\end{equation}

The second item in the equation (6) is fees arising from such delivery at redemption.

It should be noted that the issuer may agree in advance that when redeeming RSDM collateral, the quantity must be an integer multiple of a certain minimum value, such as the minimum unit being kilograms. Of course, RSDM holders can also sell RSDM to others or make purchases, just like how people hold stablecoins.

In a sense, RSDM is a monetized warehouse receipt. However, it is different from ordinary warehouse receipts, which cannot be circulated as currency. RSDM is also different from traditional tokens such as gold certificates, as the collateral assets of tokens are usually used for risky investments. In such cases, if the issuer fails in its investments, RSDM holders may be unable to redeem the collateral. Of course, RSDM, like fiat currency, can be deposited in banks, which can use it as an anchor for currency issuance or credit creation.

The collateral for RSDM must be entirely stored by a third-party professional custodian and cannot be used for any risky investments. The storage cost of collateral is offset by the depreciation of RSDM's face value. It is precisely this rule of face value reduction based on negative interest rates that lays the foundation for a 100\% convertible modern honest currency.

\subsection{The “good” collateral for the anchor of modern honest money}
In the above discussion, we did not analyze the attributes of RSDM's collateral, but generally believed that precious metals were a better choice. In fact, as a safe haven of personal wealth, the RSDM that people need is such that when they are redeemed, a middle-class family has the conditions to safely store these collateral. For example, an ordinary household safe can easily store one kilogram of gold. If the collateral of an RSDM is steel, and the total residual value of the RSDM held by someone is 200 tons of steel, when these RSDMs are redeemed, ordinary households generally do not have the conditions to store 200 tons of steel. In fact, it is also difficult for ordinary households to store 10 tons of copper. However, storing 80 kilograms of silver may have this condition. Obviously, most households have the condition to store one kilogram of gold. Although the value of these collateral (200 tons of steel, 10 tons of copper, 80 kilograms of silver, 1 kilogram of gold) is almost on the same level of magnitude.

According to such standards, precious metals can be considered a good collateral for honest money. RSDM, which uses the “good” collateral assets as currency anchors, can be called \textit{\textbf{modern honest currency}}.

\subsection{Comparison of RSDM and Stablecoin/RWA}
RSDM is essentially a symbolized commodity money. In order to compensate for increasing storage costs of collateral over time, the denomination of RSDM automatically depreciates at a negative interest rate, which is the cornerstone of the 100\% convertibility of commodity money tokens.

In theory, all money can be viewed as commodities, and similarly, all commodities and assets also have the possibility of monetization. In this sense, RSDM, stablecoin and RWA can all be seen as a new type of currency. From the perspective of asset securitization, in a broad sense, RWA is just a type of stock with different issuance conditions and trading venues compared to traditional stocks. Obviously, stocks cannot be redeemed for the tangible assets they represent and can only be traded. Most RWAs are also unable to redeem the corresponding assets. For example, you cannot redeem one tenth of a house. 

Commodity-collateralized stablecoins(CCS), especially those with precious metals as reserve assets, have a value that is linked to commodity prices. For example, gold-collateralized stablecoins (GCS), whose collateral is generally stored in authoritative certified vaults (such as LBMA certified vaults in London and Zurich vaults in Switzerland), support fragmented trading and can be redeemed for physical gold, of course, a minimum redemption amount must be met. It seems that GCS has both the function of value storage and the advantage of low transaction costs of digital currency. The holder of the token needs to pay a delivery fee when redeeming gold. These features are almost identical to RSDM. The difference is that the denomination of GCS does not have a self-decaying mechanism. Obviously, the third section of this paper has proven that as long as the issuer only earns profits through the minting and redemption fees of the token, and these fees are unrelated to the duration of the token's existence, there will come a day when the issuer will be unable to sustain operating costs, leading to bankruptcy and interruption of redeemability.

In fact, commodity-collateralized stablecoins were an early practice of RWA, which made people understand that many physical assets can be efficiently circulated through blockchain technology. It is not difficult to see that RWA provides broader asset support for stablecoins, and in the future, stablecoins can explore using more types of RWA as collateral. The differences in the main attributes of RSDM, CCS and RWA tokens are shown in Table 1.

\begin{table}[htbp]
    \centering
    \small
    \renewcommand{\thetable}{1}
    \caption{\textbf{Comparison of main attributes of RSDM, CCS and RWA}}
    \newcolumntype{Y}{>{\centering\arraybackslash}X} 
    \begin{tabularx}{\textwidth}{c Y Y Y Y}
        \toprule
        & \makecell{Promise\\ redeemable} & \makecell{Redeemable\\ attributes} & \makecell{Mortgage\\ asset risk} & \makecell{Dynamic or static\\ storage fees} \\
        \midrule
        RSDM & $\checkmark$ & $\checkmark$ & low & Dynamic \\
        CCS & $\checkmark$ & $\times$ & high & static \\
        RWA & Partial & $\times$ & high & static \\
        \bottomrule
    \end{tabularx}
    \label{tab:comparison}
\end{table}

It is not difficult to see from Table 1 that as a safe haven for personal wealth, it is a wise choice for people to moderately hold RSDM with precious metals as collateral. In terms of long-term stability of purchasing power, RSDM is superior to the average value of fiat currency. Therefore, RSDM can be better used for long-term contract settlement in commodity trade, avoiding losses for buyers or sellers due to the severe depreciation of certain fiat currencies.

\section{Building the multi-monetary system based on RSDM: guarding the reverse Gresham’s law}
In the current era of credit currency, almost all fiat currency circulating in various regions of the world is paper money. The history of human money is basically a process of artificially creating inflation, and the only difference is that countries control the depreciation speed of their own fiat currencies differently. Although people can store wealth by storing precious metals, the implicit storage cost is very high. Therefore, how to provide people with a good wealth storage tool has always been an important content in the history of money evolution.
\subsection{The risk of hyperinflation in the rising wave of over issuance of fiat and token money such as stablecoins/RWA.}
Issuing more cheap currencies without intrinsic value has always been a common trend in the international community, and the legalization of stablecoins will strengthen this trend as it helps to enhance the government's ability to expand fiscal deficits. The issuance of stablecoin, especially fiat-collateralized stablecoins, will inevitably lead to credit creation, even if it is a 1:1 pegged dollar. Take USDC as an example, when people buy USDC, the issuer will use the received dollars to buy US treasury bond bonds. After receiving funding, the US Treasury Department expanded its bond issuance quantity, and the new US dollar flowed globally. This cycle causes the liquidity of the US dollar to snowball, and from this perspective, stablecoins are a transitional path for currency over issuance. Compared to traditional currencies, according to Fisher's formula, the high circulation efficiency of stablecoins will amplify purchasing power.

The emergence of Real World Asset Tokenization has endowed some commodities with the attribute of “money”. This indicates that all physical and virtual assets in the world, including securitized assets, can be monetized. Due to the fact that RWA can be used as a “money” to undertake payment and settlement functions, it will further affect the existing monetary system.

When the market size of stablecoins and RWAs reaches the order of magnitude of the global nominal GDP, coupled with the continuous depreciation of existing fiat currencies, it will inevitably lead to excessive currency chasing relatively few goods and services, and prices will be a wild horse out of control, and people's wealth will continue to evaporate.
\subsection{It is difficult for a single currency to shoulder the responsibility of modern ‘good’ money}
For any fiat currency, the government often has the impulse to issue excess currency. Observing over a longer period of time, the overall trend of fiat currencies is depreciation because they lack the support of intrinsic value. Fiat currency has almost no long-term value storage function.

Commodity money represented by gold and silver, although has a good wealth storage function, its disadvantages are also obvious: (1) The supply cannot meet the demand for economic growth; (2) Lack of flexibility in responding to economic crises; (3) The logistics cost of circulation and storage is high; (4) The cost of identifying authenticity, testing purity, and verifying weight is high.

Virtual currencies such as Bitcoin, due to their high price volatility, are clearly not suitable for long-term contract settlement. However, fiat-collateralized stablecoins can be seen as shadow currency of fiat currency, with no essential difference in attributes from fiat currency.

As for RSDM, its main disadvantage is limited resources. RSDM anchored on a single commodity usually cannot meet the growth of GDP in terms of volume.

From the above analysis, it can be seen that a single type of currency is difficult to perfectly achieve the three basic functions required of modern good money: store of value, low logistics costs for circulation, and the ability to flexibly adjust the issuance scale to meet changes in GDP.

For example, a sole gold standard is difficult to form a perfect monetary system. 
\citet{baur2010}, through long-term empirical analysis, revealed that gold's annualized volatility (typically exceeding major reserve currencies) introduces inherent risks when used as a sole monetary anchor. Price fluctuations in gold could destabilize economies reliant on such a system.

\subsection{The advantages of the multi-monetary system consisting of fiat/CBDC and RSDM}
According to \citet{hayek1990}, in a multi-monetary system, the demand for a certain currency depends on its purchasing power trend, specifically, on its depreciation or appreciation relative to a basket of currencies. In a free market of currency, people are always ready to sell their currency and buy other currencies, and they keep their good money in their hands.

If we view the world as a country where the government allows private banks to issue currency, people are willing to choose the currency with the slowest depreciation. If all credit currencies are severely over issuance, people will have no choice but to preserve physical commodities such as precious metals, although the storage and opportunity costs of doing so are high.

In the era of credit currency, there is no low-cost safe haven for private wealth. Therefore, only by introducing RSDM based on precious metals can the multi-monetary system become more valuable. RSDM with gold or silver as collateral is like a catfish in the catfish effect, which can deter the bottomless depreciation of credit currencies. RSDM can serve as a “anti-retreat troops” to prevent the depreciation of fiat currency.

As a modern ``good'' money under the Internet environment, it must have two basic functions: long-term value storage and zero logistics cost of currency circulation. Therefore, only parallel monetary systems including RSDM, such as a triple- monetary system consisting of RSDM, domestic fiat and major international reserve currencies, can form ultimate safe haven for wealth and safeguard the reverse Gresham law.

\section{Optimization selection model for currency types in a multi-monetary pool}

Obviously, it is more convenient to price a product in a single money than in multiple money. Sellers usually consider using multiple currencies for pricing only when absolutely necessary. For example, under silver standard system, when copper coins are used as secondary money, if the government sets the exchange ratio between silver and copper, there may be certain periods of time when unscrupulous merchants melt copper coins for profit. Using multiple currencies to price goods can easily lead to price chaos and debt disorder. Due to the multitude of potential currency types, selecting which currencies to form a good monetary system requires optimization in alternative currency combinations. For this purpose, in this section, we will develop a general mathematical model to solve the optimization selection problem of a multi-monetary system.

\subsection{A complete set of monetary functions that a ``good'' money should have under the Internet environment}

The classic functions of traditional money, namely unit of account (measure of value), medium of exchange, store of value, and means of payment, have gradually formed with the development of commodity economy. Among them, the unit of account and medium of exchange are the most fundamental functions of money. A good modern money not only needs to retain these basic functions, but also needs to consider the logistics cost, anti-counterfeiting performance, and anti-hoarding performance of the money, and so on. If we take these additional properties that a good modern money must possess as the extended functions of money, they can form a complete set of monetary functions. 

\begin{equation}
F^{\text{Money}} = \left\{ F_1^{\text{Money}}, F_2^{\text{Money}}, ..., F_{N^{\text{Function}}}^{\text{Money}} \right\}
\label{eq:7}
\end{equation}

\noindent where, $N^{\text{Function}}$ represents the number of functions that a ``good'' money should possess. For example,

$F_1^{\text{Money}}$= unit of account, money of account, measure of value.

$F_2^{\text{Money}}$= medium of exchange.

$F_3^{\text{Money}}$= means of payment

$F_4^{\text{Money}}$= store of value

$F_5^{\text{Money}}$= Can prevent hoarding.

$F_6^{\text{Money}}$= Low logistics (circulation and storage) costs.

$F_7^{\text{Money}}$= No weight loss during circulation.

$F_8^{\text{Money}}$= There is a possibility of matching the money supply with wealth (such as GDP).

$F_9^{\text{Money}}$= Having stable purchasing power to meet long-term contract needs.

$F_{10}^{\text{Money}}$= Can be used as a money for taxation.

$F_{11}^{\text{Money}}$= It's difficult to over issue.

$F_{12}^{\text{Money}}$= Low anti-counterfeiting cost and easy identification of counterfeit and inferior money.

\subsection{The set of ``general equivalents'': Signs of value as a special commodity}

Money originated from commodity exchange and is essentially a special commodity that serves as a fixed general equivalent. According to modern monetary theory, currency is just a unit of account. In this sense, money is a sign of value for all commodities and does not necessarily have intrinsic value, provided that there is strict discipline to constrain the money supply. Following this approach, we can consider both credit and virtual currency as commodities, similar to a financial assets or a digital artwork, although they are only symbols of value. So, building a good monetary system is to search for one or several special commodities as currency in the set of commodities defined by equation (7) below. 

We denote the set of general equivalents as $S^{\text{Equivalents}}$, which consists of the following:

\begin{equation}
\begin{aligned}
S^{\text{Equivalents}}
&= \left\{ C_1^{\text{Currency}},...,C_\kappa^{\text{Currency}} \right\} \\
&= S^{\text{Fiat}} \cup S^{\text{Commodity}} \cup S^{\text{Crypto}} \cup S^{\text{RSDM}} \cup ,\dots, \cup S^{\text{Other}}
\end{aligned}
\label{eq:8}
\end{equation}

\noindent where, $C_k^{\text{Currency}}$ represents the $k$-th currency or sign of value or monetizable commodity, $S^{\text{Fiat}}$ is the set of various credit and fiat currencies, $S^{\text{Commodity}}$ represents the set of commodity money, $S^{\text{Crypto}}$ is the set of virtual currencies, $S^{\text{RSDM}}$ is the set of redeemable self-decaying money, and $S^{\text{Other}}$ represents the set of other forms of assets or currencies, such as some tokenized RWAs. The expansion form of each subset in equation (8) is as follows:

\begin{equation}
S^{\text{Fiat}} = \left\{ T_1^{\text{Fiat}}, T_2^{\text{Fiat}}, ..., T_{N^{\text{Fiat}}}^{\text{Fiat}} \right\}
\label{eq:9}
\end{equation}

\begin{equation}
S^{\text{Commodity}} = \left\{ S_1^{\text{Commodity}}, S_2^{\text{Commodity}}, ..., S_{N^{\text{Commodity}}}^{\text{Commodity}} \right\}
\label{eq:10}
\end{equation}

\begin{equation}
S^{\text{Crypto}} = \left\{ T_1^{\text{Crypto}}, T_2^{\text{Crypto}}, ..., T_{N^{\text{Crypto}}}^{\text{Crypto}} \right\}
\label{eq:11}
\end{equation}

\begin{equation}
S^{\text{RSDM}} = \left\{ T_1^{\text{RSDM}}, T_2^{\text{RSDM}}, ..., T_{N^{\text{RSDM}}}^{\text{RSDM}} \right\}
\label{eq:12}
\end{equation}

\begin{equation}
S^{\text{Other}} = \left\{ T_1^{\text{Other}}, T_2^{\text{Other}}, ..., T_{N^{\text{Other}}}^{\text{Other}} \right\}
\label{eq:13}
\end{equation}

In Equation (9)--(13), $N^{\text{Fiat}}$, $N^{\text{Commodity}}$, $N^{\text{Crypto}}$, $N^{\text{RSDM}}$ and $N^{\text{Other}}$ represent the number of various types of fiat currency, commodity money, virtual currency, RSDM and tokenized RWAs, respectively. For example,

\begin{equation}
S^{\text{Fiat}} = \left\{ \text{USD, EUR, CNY, JPY, GBP, CAD},..., \text{SDR} \right\}
\label{eq:14}
\end{equation}

\begin{equation}
\begin{aligned}
S^{\text{Commodity}} &= \left\{ \text{Gold, Silver, Copper,} \right. \\
&\quad \left. \text{Commodity-1, Commodity-2},..., \text{ Commodity-}n \right\}
\end{aligned}
\label{eq:15}
\end{equation}

\begin{equation}
\begin{aligned}
S^{\text{Crypto}} &= \left\{ \text{BTC,ETH,XRP,DOT,SOL},..., \text{USDT,USDC}, \right. \\
&\quad \left. \text{BUSD}, \text{USDD,TUSD,XAUT} \right\}
\end{aligned}
\label{eq:16}
\end{equation}

\begin{equation}
\begin{aligned}
S^{\text{RSDM}} &= \left\{ \text{RSDM}_{\textit{Gold}}, \text{RSDM}_{\textit{Silver}},..., \right. \\
&\quad \left. \text{RSDM}_{\textit{Commodity-}\textup{1}},..., \text{RSDM}_{\textit{Commodity-n}} \right\}
\end{aligned}
\label{eq:17}
\end{equation}

\begin{equation}
S^{\text{Other}} = \left\{ \text{RWA}_1, \text{RWA}_2, ..., \text{RWA}_n \right\}
\label{eq:18}
\end{equation}

Equation (8) and its expansion well illustrate the consensus that all money is a special commodity that serves as a general equivalent or a universally recognized signs of value. Commodity money, RSDM, credit currency, virtual currency, and RWAs are convergence at the semiotic level\citep{vasantkumar2019}.

\subsection{A 0-1 integer programming model for optimizing multi-monetary system}

Note that not every currency satisfies all the functions in the complete set of monetary functions. Therefore, an excellent monetary system should be able to cover all these functions while containing as few types of currency as possible. If no money can satisfy all the functions in the set $F^{\text{Money}}$, how to build a multi-monetary system to achieve all the functions of excellent modern currency? In a sense, monetary history is the process of unifying multiple currencies into fewer or single currency. Partially drawing on this historical experience, from an optimization perspective, we believe that an excellent monetary system should be able to cover all the functions of the “good” money while containing as few types of currency as possible. For this, we will propose a linear binary programming model for optimizing Multi-monetary System Problem(MSP). Some analysis of the model will also be discussed in this section.
The parameters, sets and decision variables used in this section are listed in Table 2. 

\begin{table}[htbp]
    \centering
     \small
    \renewcommand{\thetable}{2}
    \caption{\textbf{The definitions of parameters, sets, and variable}}
    \begin{tabularx}{\textwidth}{lX}  
        \toprule
        Symbol & Definition \\
        \midrule
        $S^{\text{Mandatory}}$ & The set of the money that must serve as circulating currency in a monetary system. \\
        $u_{ck}$ & The proportional value (score) of the $k$-th function of money covered by currency C, which is a real number between 0 and 1. \\
        $w_k$ & Importance weight of the $k$-th monetary function. \\
        $N^{\text{Parallel}}$ & The number of currency types allowed to circulate in a monetary system. \\
        $H_k^{\text{Threshold}}$ & The threshold for the total share of the $k$-th monetary function possessed by all types of currencies in a monetary system. \\
        $\beta^{\text{Balance}}$ & Parameters for balancing the number of currency types. \\
        $x_c$ & 0-1 decision variables, if $x_c=1$, it means that currency C is included in the monetary system. \\
        \bottomrule
    \end{tabularx}
    \label{tab:parameters}
\end{table}

For a country or region's monetary system, only one type of currency is optimal from the perspective of the complexity of product pricing.  Therefore, one of the goals of optimizing MSP is to select as few money types as possible in the monetary system. On the other hand, the number of selected money types should cover as many elements as possible in the set $F^{\text{Money}}$. Based on this goal and using above-mentioned notations, the MSP can now be written with a 0-1 integer programming formulation as follows:

\begin{equation}
\begin{aligned}
\max Z
&= \sum\limits_{c \in S^{\text{Equivalents}}} \Bigl( x_c \sum\limits_{k \in F^{\text{Money}}} w_k u_{ck} \Bigr)
 \;-\; \beta^{\text{Balance}} \sum\limits_{c \in S^{\text{Equivalents}}} x_c
\end{aligned}
\label{eq:19}
\end{equation}

\noindent (MSP)\quad S.t.

\begin{equation}
\sum\limits_{c \in S^{\text{Equivalents}}} x_c \;\le\; N^{\text{Parallel}}
\label{eq:20}
\end{equation}

\begin{equation}
\sum\limits_{c \in S^{\text{Equivalents}}} u_{ck}\, x_c \;\ge\; H_k^{\text{Threshold}},\quad \forall\, k \in F^{\text{Money}}
\label{eq:21}
\end{equation}

\begin{equation}
x_c = 1,\quad \forall\, c \in S^{\text{Mandatory}}
\label{eq:22}
\end{equation}

\begin{equation}
x_c \in \{0,1\},\quad \forall\, c \in S^{\text{Equivalents}}
\label{eq:23}
\end{equation}

In the objective function of model MSP, the first term is the sum of scores for various monetary functions. And the second term of the objective function is to minimize the number of selected currencies of various types. Constraint (20) is to limit the number of currency types allowed to circulate in parallel in the monetary system, in order to avoid confusion in commodity pricing caused by overly complex currency composition. Constraint (21) is to ensure that each monetary function requires a minimum score. Constraint (22) requires that a certain currency must be included in the monetary system, such as the domestic fiat currency. Finally, Constraint (23) is the binary constraints on decision variables.

For ease of calculation, we quantify each monetary function as a real number between 0 and 1. It is not difficult to find that the cumulative sum of the first term in equation (19) may be greater than 1. To avoid this situation, we can modify the objective function of MSP to the following form:

\begin{equation}
\begin{aligned}
\max Z
&= \sum\limits_{k \in F^{\text{Money}}}
   \min\!\left\{
      1,\;
      \sum\limits_{c \in S^{\text{Equivalents}}} w_k u_{ck} x_c
   \right\}
   \;-\; \beta^{\text{Balance}} \sum\limits_{c \in S^{\text{Equivalents}}} x_c
\end{aligned}
\label{eq:24}
\end{equation}

Obviously, after the objective function is reconstructed, the MSP model will become a nonlinear integer programming model, and the calculation of taking the minimum among the set elements is non differentiable.

\subsection{Several inspirations about MSP model}
Let $A_k^{\text{Function}}$ be the set of monetary functions possessed by the $k$-th type of currency $C_k^{\text{Currency}}$. Generally, the condition $A_k^{\text{Function}} \subset F^{\text{Money}}$ is met. If there exists currency $C_*^{\text{Currency}}$ that satisfies the condition $A_*^{\text{Function}} = F^{\text{Money}}$, we believe that this currency belongs to a `good' money.

\textbf{Situation 1}: $N^{\text{Parallel}}=1$, which means that only one money can circulate in the monetary system, which is currently the situation in most countries. In this case, the possible choices for the standard money are: domestic fiat currency, fiat currency of other countries/regions, RSDM, Virtual currency.

For most countries' fiat or stablecoin anchored in fiat currency, that is, $C_k^{\text{Currency}} \in S^{\text{Fiat}}$, their functions are usually difficult to cover $F_4^{\text{Money}}$, $F_9^{\text{Money}}$, $F_{11}^{\text{Money}}$. Of course, some fiat currencies with relatively strict issuance discipline, such as the US dollar, euro, etc., can approximately satisfy $F_9^{\text{Money}}$ during certain periods.

If a physical commodity money, $C_k^{\text{Currency}} \in S^{\text{Commodity}}$, is chosen as the standard currency, its functions cannot satisfy $F_5^{\text{Money}}$, $F_6^{\text{Money}}$, $F_7^{\text{Money}}$, $F_8^{\text{Money}}$, $F_{12}^{\text{Money}}$. Especially in the early stages of circulation, their volatility will be significant, although over time they will enter a stable state of purchasing power.

If virtual currencies or tokenized RWAs are chosen, they are difficult to satisfy $F_1^{\text{Money}}$, $F_4^{\text{Money}}$, $F_5^{\text{Money}}$, $F_8^{\text{Money}}$, $F_9^{\text{Money}}$, $F_{10}^{\text{Money}}$, $F_{13}^{\text{Money}}$ and $F_{14}^{\text{Money}}$. Although some virtual currencies such as Bitcoin are limited in their total issuance by algorithms, similar cryptocurrencies can have an infinite number, so overall they do not meet $F_8^{\text{Money}}$ and $F_9^{\text{Money}}$ requirements.

If a certain RSDM is chosen as the standard currency, such as RSDM with gold as collateral, the volume of a single commodity is difficult to meet $F_8^{\text{Money}}$.

According to the above analysis, any single type of currency is usually unable to meet the standards of modern ``good'' money. 

\textbf{Situation 2}: $N^{\text{Parallel}}>1$.

In this case, some countries may have two or more currencies circulating simultaneously. For example, Argentina's ``bi monetary economies'' policy allows daily necessities and commodities to be priced in both pesos and US dollars. From a long-term perspective, even a multi-monetary system with multiple fiat currencies circulating in parallel usually cannot meet all the functions of a ``good money'' in the digital age, because the common problem of all fiat currencies is endless depreciation, only the speed of depreciation is different, and they cannot achieve long-term value storage.

It is not difficult to find that a bi-monetary system consisting of fiat currency with strict issuance discipline and RSDM anchored with precious metals can cover all the functions of ``good'' money. For example, in the Eurozone, if RSDM is allowed to circulate in addition to the Euro, the following equation clearly holds:

\begin{equation}
A_{\mathit{EUR}}^{\text{Function}} \cup A_{\mathit{RSDM}}^{\text{Function}} \supseteq F^{\text{Money}}
\label{eq:25}
\end{equation}

When the Euro experiences a severe depreciation (although this probability is very low), the market will automatically choose RSDM as the money of account or as the measure of value for goods. If the European Central Bank continues to over issue euros, it will accelerate the rate at which the euro is abandoned by the market.

Generally speaking, any monetary system with multiple currencies running in parallel, as long as their combined functions can satisfy all the functions of a ``good money'', can constitute a stable price measurement standard. From a historical perspective, it was only under the gold standard from the early 18th century to the early 20th century that the government's excessive issuance of currency was constrained by precious metal resources, and the purchasing power of money stabilized over a longer period of time.

From the evolution of money, it can be inferred that a triple monetary system consisting of domestic fiat currency, a major international settlement currency, and a precious metal anchored RSDM can better meet all the functions of a ``good money'' and achieve the goals of long-term wealth storage and price stability.

\section{Potential application scenarios of RSDM in the real world}

Given that RSDM has strict physical reserve supervision and a commitment to ensure redemption at any time, it is an honest currency that is difficult to overissue. In this section, we will discuss several typical application scenarios of RSDM, including a new approach to activate dormant gold assets in India based on RSDM, and the gold monetization scheme in the United States, and a triple-monetary system based on RSDM in countries with high inflation.

\subsection{A new way to activate India's dormant gold assets: RSDM based gold monetization scheme}
\subsubsection{Introduction to India's Gold Monetization Scheme}
The India's Gold Monetization Scheme began in 2015, aiming to revitalize the approximately 25000 tons of gold reserves in the private sector. The main purpose of this plan is to encourage people to deposit their idle gold in banks, which will pay interest to depositors based on the price of gold in rupees (only slightly over 2\%) on the day of the transaction. After maturity, depositors can choose to receive an equal amount of gold or cash. The Indian government will auction or lease melted gold to jewelers to reduce import demand. The Bank of India's SGB is more like a financial experiment without hedging.

In fact, long ago, some major Indian gold jewelry merchants offered very attractive annualized returns to customers who lent gold, in order to attract people to lend their idle gold.

\subsubsection{The main shortcomings of India's gold monetization scheme}

India's gold monetization scheme failed to achieve the expected results, mainly due to the following reasons:

\begin{enumerate}[label=(\arabic*)]
    \item In the initial stage, the distribution of gold testing centers and storage centers could not meet customer needs, making it difficult for depositors to easily complete the process of depositing gold into the bank, seriously affecting their willingness to participate. Although the government increased purity testing centers in the later stage, a negative impression had already formed in the early stage, leading to poor results in subsequent investment.
    
    \item The gold deposit interest rate is much lower than the cash deposit interest rate, and depositors need to bear the cost of melting gold to test its purity and the loss of some gold, resulting in lower actual returns.
    
    \item Gold has a profound emotional value in Indian culture, and the general public is reluctant to melt and deposit gold in banks.
    
    \item The policy design did not arouse the interest of high-net-worth individuals (accounting for 85\%--90\% of gold consumption).
\end{enumerate}

\subsubsection{A new approach to monetizing idle gold in India based on RSDM}

According to the concept of redeemable self-devalued money, a feasible way to activate dormant gold assets in India based on RSDM is expressed as follows:

\begin{enumerate}[label=(\arabic*)]
    \item Dispel residents' concerns that gold certificates cannot be redeemed for gold at any time. Allow international and domestic financial institutions to issue gold RSDMs and establish exchange points in India. Under this premise, the Indian government can also issue RSDM, and various gold RSDMs that meet regulatory requirements can circulate in parallel in the market.
    
    \item The Indian government regards RSDM as a currency circulating parallel to the rupee, and may, if necessary, designate RSDM as a tax money, allowing goods to be priced in both rupees and gold RSDMs.
    
    \item Allow banks to operate deposit and loan business for gold RSDM. RSDM holders can deposit RSDM in the bank in exchange for interest. For example, if the decay coefficient of RSDM is annualized to -2\%, the interest paid by the bank in gold is 3\%, and the actual benefit to depositors is annualized to 1\%. Of course, the interest can also be in rupees, which depends on the preferences of financial institutions and market demand.
    
    \item In the context of RSDM as a currency circulation, people can exchange gold jewelry for RSDM at RSDM sales outlets (customers need to bear the weight loss of gold melting and purity testing fees). Residents can also purchase RSDM in rupees, and vice versa. This two-way exchange is not time limited, just like foreign exchange trading.
\end{enumerate}

After converting physical gold into digital or paper RSDM, if there is a positive return, considering the storage costs and security risks of gold in households or temples, gold holders with the purpose of value storage (especially those high net worth individuals) should be willing to accept it.

\subsection{RSDM as the ultimate competition with the US dollar will become a soft constraint to prevent the over-issuance of fiat currency}

\subsubsection{The crisis and challenges faced by the US dollar system}

Over the past century, the purchasing power of the US dollar has depreciated by an average of about 3\% per year. Compared to nearly two hundred sovereign currencies in the world, the decline in purchasing power of the US dollar is not significant, making it one of the least devalued good currencies.

However, since the decoupling of the dollar and gold in 1971, its depreciation rate seems to have accelerated and is facing significant downward pressure. It may be at the beginning of a long-term depreciation. At present, the US fiscal deficit and debt scale continue to expand, and can only be sustained through the ``borrowing new to repay old'' model. The credit of the US dollar is gradually being consumed by some short-sighted actions.

The rising tide of virtual currencies such as stablecoins is stimulating countries to accelerate the launch of their own digital currencies, which may lead to a more diversified monetary system. More and more bilateral and regional trade is ``bypassing'' the US dollar system, and many countries are promoting local currency settlement (LCS). The global trend of ``de-dollarization'' is spreading, which may reduce the demand for US dollars in the long run. In addition, the new energy revolution is in full swing, making it difficult for the ``petrodollar'' to regain its former glory. Therefore, the US dollar is also forced to start searching for new anchors.

In this context, many states in the United States hope to establish a multi-monetary system at the state level by recognizing gold and silver as legal tender, providing the public with tools to resist the depreciation of the US dollar. On June 30, 2025, Texas Governor Greg Abbott signed a law officially recognizing gold and silver specie as legal tender, granting Texans the right to use physical gold and silver (coins or bars) in daily transactions, including commercial trade, debt repayment, and private contracts.

Florida Governor Ron DeSantis signed the minHB 999 bill, recognizing gold and silver as the state's legal tender. The bill will come into effect on July 1, 2026.

In recent decades, the Federal Government Mint has minted and issued hundreds of millions of ounces of gold and silver eagle coins. Therefore, the use of gold and silver coins as legal tender in various states of the United States has a good market foundation.

\subsubsection{RSDM based gold and silver monetization pathway in the United States}

Although the implementation details of the minHB 999 bill have not yet been released (to be formulated before November 1, 2025), it can be concluded that physical gold and silver coins are difficult to circulate comprehensively due to their inherent defects, such as the high cost of testing the purity of gold (transactions need to be completed through electronic custody systems).

If physical gold and silver coins are directly used for transactions, it will involve a series of troublesome issues, such as identifying counterfeit coins, purity testing, insufficient weight due to wear and tear, and finding change for small transactions. The current practice is that if a customer wants to purchase a product priced at 0.3 silver eagle coins. Users must first deposit their Silver Eagle coins into a silver custody account recognized by the state government. When customers make purchases, they calculate the silver price of the goods through the custody system, generate an electronic silver check, send the QR code or URL to the store cashier, and the store verifies the legitimacy of the check and confirms the transfer to the account. After the payment was completed, the customer's silver balance decreased by 0.3 ounces, and the merchant's account increased by 0.3 ounces of silver. Merchants can keep these silver coins, or exchange them into US dollars, or directly withdraw physical silver eagle coins after accumulating a certain amount.

If the government recycles physical gold and silver coins and replaces them with tokens, promising that the tokens are 100\% convertible, how can it avoid hoarding by residents, and how can state governments or the US Treasury bear long-term gold and silver storage costs? Based on the previous analysis, it is evident that there will be a black hole in storage costs, unless the government uses tax subsidies to cover the storage expenses of precious metals. In fact, over a hundred years ago, the United States issued convertible Silver Certificates.

Based on the previous analysis, it is not difficult to find that as long as physical gold and silver coins are converted into RSDM, the circulation of gold and silver as legal tender becomes simpler.

The issuer (state government, federal mint, or other financial institution) can issue gold or silver RSDMs of different weights to ensure that they can replace the US dollar in daily transactions, including commercial trade, debt repayment, and private contracts. The value of the attenuation coefficient should consider both the logistics costs of storing and transporting gold or silver, as well as ensuring that it is not hoarded by most residents for a long time.

If the storage cost of gold or silver is entirely borne by the holders of RSDM, the design of the attenuation coefficient should balance the depreciation rate of the US dollar. Of course, RSDMs issued by financial institutions can be mainly based on centralized custody accounts, or partially adopt blockchain technology, such as referencing the technical framework of stablecoins. RSDMs should be traceable to the corresponding gold or silver collateral, and certain specifications of RSDM can be redeemed for gold and silver eagle coins. All RSDMs can be exchanged for physical gold or silver to ensure market trust.

In theory, RSDMs issued by any compliant financial institution around the world can circulate in the United States. Of course, RSDMs issued by the US Mint or other US financial institutions can also circulate internationally on an equal footing, with the requirement that the establishment of exchange points be synchronized with internationalization.

\subsection{Triple-monetary system for high inflation countries based on RSDM}

Although the purchasing power of mainstream international reserve currencies such as the US dollar, euro, and pound is relatively stable at present, many countries are still in a state of hyperinflation. For example, due to the severe depreciation of the Zimbabwean dollar, the Zimbabwean government once abolished ZWD and adopted a new monetary system of nine currencies in parallel, including the US dollar and the euro. Later, the country issued “bond currencies” and “Zimbabwe Gold”, but both were not effective due to lack of market trust.

In the past few decades, Argentina has been in a state of hyperinflation, and since March 1, 2025, the government has implemented an “bi-monetary economies” policy. Merchants can legally accept payments in US dollars without the need to exchange them for pesos.

In fact, for these countries with high inflation, the best way to deter currency depreciation is to establish a triple-monetary system: domestic fiat currency, a representative international settlement currency (such as the US dollar or euro), and precious metal RSDM issued by reputable international financial institutions. That is, these three types of currencies circulate in parallel. Under this triple-monetary system, it is possible to fully activate the private sector's dollars and gold, allowing them to re-enter the economic cycle, thereby avoiding the situation of a large number of “dollars under the mattress” and “gold in the jewelry box”, and indirectly constrain the rapid depreciation of the domestic fiat currency. 

\section{Analysis of RSDM demand with precious metals as collateral}

The money supply refers to the total amount of currencies available for transactions and payments in an economic system. At present, besides fiat currencies, there are also virtual currencies and RWAs that can be used for transactions and payments.
Obviously, virtual currency and tokens such as RWA and RSDM will all affect the total money supply. Considering that fiat collateralized stablecoins are essentially ``shadow tokens'' of fiat currencies, we classify them as the fiat currency system, and classify crypto-collateralized Stablecoins and algorithmic stablecoins as virtual currencies. The stablecoin mentioned later refers to a fiat collateralized stablecoin, unless otherwise specified.
Here, we consider all economic activities in the world (which can also be a country or a region) as an economic system, and the total money supply of this system is denoted as $M^{\text{Supply}}$. Its sub components can be expressed as follows:

\begin{equation}
\begin{aligned}
M^{\text{Supply}}
&= \kappa^{\text{Muiltiplier-fiat}} B^{\text{Reserve-fiat}} \\
&\quad + \kappa^{\text{Muiltiplier-SDM}} B^{\text{Reserve-SDM}} + C^{\text{Other}}
\end{aligned}
\label{eq:26}
\end{equation}

In the formula, $B^{\text{Reserve-fiat}}$ is the total amount of base money corresponding to fiat currencies, and $B^{\text{Reserve-SDM}}$ is the amount of RSDM with gold (or other commodities) as collateral. $\kappa^{\text{Muiltiplier-fiat}}$ is the average money multiplier of various fiat currencies, and $\kappa^{\text{Muiltiplier-SDM}}$ is the money multiplier of RSDM. $C^{\text{Other}}$ is the money supply of various virtual currencies (excluding Fiat collateralized stablecoins) and RWAs converted into US dollars or gold.
It should be pointed out that from the subscription and redemption mechanism of some stablecoins and their 100\% asset reserve characteristics, the stablecoins issued by issuing institutions are similar to M0 in the real world and are the base money on the chain; And stablecoins in circulation, due to financial instruments such as on chain lending and pledging, have the function of credit creation, which can be compared to M2 in the real world.

Therefore, stablecoins have some attributes of M0 and M2. Here, we absorb the impact of stablecoins on the money supply by increasing the money multiplier $\kappa^{\text{Muiltiplier-fiat}}$.
The relationship between broad money and gross domestic product is (here analyzed from a global perspective, ignoring the difference between GDP and GNP):

\begin{equation}
M_{2} = M^{\text{Demand}} = K^{\text{Marshallian}} \times V^{\text{GDP}}
\label{eq:27}
\end{equation}

\noindent where, $K^{\text{Marshallian}}$ is the monetization rate, also known as the Marshallian K. Considering the equilibrium equation $M^{\text{Supply}} = M^{\text{Demand}}$, there is the following equation:

\begin{equation}
\begin{aligned}
K^{\text{Marshallian}} V^{\text{GDP}}
&= \kappa^{\text{Muiltiplier-fiat}}\, B^{\text{Reserve-fiat}} \\
&\quad + \kappa^{\text{Muiltiplier-SDM}}\, B^{\text{Reserve-SDM}}  + C^{\text{Other}}
\end{aligned}
\label{eq:28}
\end{equation}

The GDP and Marshallian K on the left side of equation (27) are usually relatively stable values, while the three items on the right side will show a trade-off relationship. Therefore, this formula better reflects Hayek's theory of parallel currency competition.

Obviously, when the supply of money outside of fiat currency increases, the demand for fiat currencies will show a decreasing trend. Therefore, currencies with long-term value storage and good liquidity, such as RSDM, will gain more and more users in competition, while anchor free currencies will fall into a vicious cycle, with more issuance, more depreciation, and fewer users. It is not difficult to infer that RSDM joining currency competition can constrain the over issuance of anchor free currencies and non-convertible tokens. Ultimately, the world economic system will stabilize in a triple parallel money system consisting of modern honest money, credit currency, and non-convertible tokens.

For example, if the Marshallian K is 0.7 and $V^{\text{GDP}}=120$ trillion US dollars, at some point in the future, assuming fiat currencies (including their corresponding stablecoins), RSDMs and other tokens each account for one-third, that is, 40 trillion US dollars. If $m^{\text{Muiltiplier-SDM}}=8.0$, then $B^{\text{Reserve-SDM}}=5$ trillion US dollars is required. When RSDM with gold as collateral is issued as currency, the role of gold will transition from commodity to money, and the price of gold is expected to have a significant increase, conservatively estimated to be above $\$5000 \sim \$10000$ per ounce. In addition, as RSDM is easier to store than physical gold, scattered gold among the public will be further concentrated in the hands of RSDM issuers. At that time, the gold used as collateral for RSDM is expected to reach a scale of over $50000$ tons. At present, the world has extracted approximately $220000$ tons of gold, with jewelry accounting for nearly half, private investment accounting for just over one-fifth, central banks holding nearly $40000$ tons, and industrial and medical use accounting for over $30000$ tons. The global annual production of gold mines is approximately $3500$ tons. Therefore, it is cautiously estimated that RSDM collateralized with precious metals can be converted to over $10–20$ trillion US dollars, which can provide more than half of the global currency demand. Considering that RSDM has a fundamental monetary attribute, the leveraged M2 will reach the order of one hundred trillion US dollars. The RSDM of these precious metals will create a crowding out effect on fiat and stablecoins, and allow more virtual currencies to enter the RWA reservoir. Of course, in order to meet the demand of some users for paper-based RSDM, we can also issue RSDMs with traditional coin metals as collateral, such as $200000$ tons of silver, $10$ million tons of copper, etc., to facilitate small-scale transactions in daily life.

At that time, RSDM will play a decisive role in maintaining the stability of global commodity prices.

\section{Conclusions}

This paper proposes the concept of modern honest money /“good money” under the Internet environment. In addition to retaining all functions of classic honest money (such as precious metal money represented by gold or silver coins), they also need to have a circulation performance of nearly zero logistics costs, similar to paper currency or digital currency. That is, they need to have intrinsic value like physical gold coin and zero logistics costs like digital currency. We further point out that tokens claiming to be able to redeem metal coin at a static face value at any time are perpetual motion machines that are impossible to achieve. Silver or gold certificates that have appeared in history and stablecoins that claim to be redeemable for gold in the world today are illusory perpetual motion machines unless they appear in the form of self-devalued denominations.

The four basic deficiencies of precious metal money in history are difficulty in identifying authenticity, difficulty in purity testing, difficulty in weight verification, and high logistics costs. Therefore, classical honest money represented by precious metals must be tokenized, and this token is essentially an electronic or paper warehouse receipt, that is, possessing 100\% collateral (such as gold or silver) and being able to redeem the collateral according to the agreed self-devalued face value. Only such token can retain the intrinsic value of classical honest money, overcome their basic deficiencies, and turn into a “good money” adapting to the Internet environment.

Since entering the era of credit currency, people have been questioning whether paper or digital value symbols, such as the US dollar or US bonds, are still a “good wealth storage tool”. Obviously, compared to the precious metal money that have circulated throughout history, they are not “good wealth storage tools”. However, in today's world, although people can store wealth by preserving physical precious metals, the implicit storage and circulation costs are high. Therefore, how to provide a good wealth storage tool for the world has always been an important content in the history of currency evolution.

According to the author's previous work, we regard RSDM as a modern honest money or a “good money” under the Internet environment. RSDM transfers part of its value by means of self-devalued face value to pay for the accumulated collateral storage costs over time. It is precisely this automatic transfer mechanism of partial value that enables the redemption of residual value of tokens and completes the transformation of commodity money symbolization.

Considering that a single type of currency is difficult to shoulder the responsibility of modern “good money”, for example, precious metal money is difficult to solve the deep contradiction between resource scarcity and global GDP growth. Therefore, this paper proposes a ternary parallel monetary system consisting of local/domestic fiat currency, major international reserve currency, and RSDM. In such a triple-monetary system, the complementary functions of different types of currency can form a stable price measurement system. Such a parallel monetary system will increase the resistance to rapid depreciation of fiat currencies or lead to the elimination of fiat currencies that depreciate uncontrollably.

In order to solve the problem of element composition in parallel monetary systems, this paper built an optimization model based on integer programming to solve how to select currency types from a set of candidate money types theoretically in order to construct the most perfect multi-monetary system. This model covers two crucial dimensions: the functional completeness of parallel monetary systems and the minimization of parallel money types. From a long-term perspective, the model suggests that a parallel monetary system composed of multiple credit currencies often cannot meet all the functions that a “good money” in the digital age should have.

As a natural application of RSDM theory, we propose a monetization scheme for Indian gold. And it is pointed out that RSDM, as the ultimate competitor with the US dollar, will become a soft constraint to prevent the over-issuance of fiat currency.

At the end of the paper, as a cautious estimate, RSDM collateralized with precious metals can reach over 10 trillion US dollars, providing half of the global currency demand.

It can be foreseen that the demand for stablecoins will significantly decrease when CBDCs can conveniently conduct peer-to-peer cross-border payments through digital wallets. RSDM, which has inherent value and almost no logistics costs, will become the protagonist of international reserve currency after the initial price fluctuation period. As an internationally recognized currency, RSDM can provide a last safe haven for private wealth and ultimately become an “automatic stabilizer” in the modern monetary system. In addition, RSDM can serve as a “supervisory currency” that serves as a tool to deter inflation and correct monetary system failures. RSDM and major international reserve currencies will eventually form a new generation of safe haven asset networks, reshaping the logic of global wealth storage and trading.

The question that needs further research is: How long is the transition period of RSDM from the role of commodity to the role of money? How much is the price fluctuation relative to a basket of commodities during this transition period? How to determine the decay coefficient of RSDM and other related issues?


\end{document}